\documentstyle[12pt]{article}
\begin{document}
\pagestyle{empty}
\vspace* {13mm}
\baselineskip=24pt
\begin{center}
%*****************TITLE****************************************
%
  {\bf GENERALIZED STATISTICS AND DYNAMICS\\ IN CURVED SPACETIME}
  \\[11mm]
  V.Bardek, S.Meljanac and A.Perica\\
  {\it Rudjer Bo\v{s}kovi\'{c}  Institute, P.O.B. 1016,\\
  41001 Zagreb, Croatia}\\
  (January 4, 1994) \\[12mm]
  {\bf Abstract}
  \end{center}
  \vskip 0.3cm
  We consider the generalized momentum-depending quon algebra in a
  dynamically evolving curved spacetime and perform a type of analysis
  similar to that of J.W.Goodison and D.J.Toms.
  We find that, at least in principle, all kinds of statistics may
  occur in some regions, i.e. phases in momentum space, depending on
  Bogoliubov coefficients determined by a specific dynamical model.
  %***************************************************************
  \newpage
  \setcounter{page}{1}
  \pagestyle{plain}
  \def\leer{\vspace{5mm}}
  \setcounter{equation}{0}%
  The problem of allowed statistics, such as Bose, Fermi, para-Bose,
  para-Fermi, anyonic, infinite-quonic, and their role in physics
  is of great importance [1-4].
  It has recently been claimed \cite{JWG} that, starting with the global
  quon algebra \cite{OWG} no generalized statistics is compatible with
  the dynamics in curved spacetime.
  The only allowed possibilities found are the ordinary Bose and Fermi
  statistics.
  However, the initial assumption that the parameter of deformation q is a real
  constant is too restrictive. Namely, two of us \cite{SM} proposed a more
  general quon algebra interpolating between all types of statistics
  simultaneously, and having the same basic properties as the global
  quon algebra \cite{OWG}.

  In this Letter, starting with the generalized quon algebra, we classify
  and discuss possible statistics compatible with gravitationally induced
  particle creation in a dynamically evolving universe.

  The generalized quon algebra \cite{SM} is given by

  \begin{equation}
  a_{i}a_{j}^{\dagger} - q_{ij}a_{j}^{\dagger}a_{i} = \delta_{ij}
  , \forall i,j\in S
  \end{equation}

  Here  $q_{ij}=q_{ji}^{*}$ are complex functions and S is a set of
  (discrete or continuos) 3-dimensional momenta. Norms in the Fock space are
  positive definite if $|q_{ij}| \leq 1$, for all pairs $i,j \in S$.
  If $\mid q_{ij}\mid < 1$, there are no commutation relations between
  $a_{i}$, $a_{j}$ (or $a_{i}^{\dagger}$, $a_{j}^{\dagger}$) operators.
  However, in the exact limit when $\mid q_{ij}\mid = 1$ for every pair of
  indices $i,j \in S$, the following commutation relations are valid:

  \begin{equation}
  a_{i}a_{j} - q_{ij}^{*}a_{j}a_{i} = 0     , \forall i,j \in S
  \end{equation}

  In this case the number operator is $N_{i} = a_{i}^{\dagger}a_{i}$.

  We consider generalized quon statistics in a dynamically evolving
  curved spacetime which, prior to some initial time $t_{1}$ and subsequent
  to some later time $t_{2}$ , is flat.The spacetime is dynamic for
  $t_{1} \leq t \leq t_{2}$ and given by the spatially flat
  Robertson-Walker metric \cite{LP}.
  In the first flat region where $t < t_{1}$ (in-region) we assume that
  annihilation and creation operators $a_{i}$, $a_{i}^{\dagger}$ satisfy
  the generalized quon algebra, Eq.(1).In the second flat region
  where $t > t_{2}$ (out-region) we assume that the corresponding
  annihilation and creation operators $b_{i}$, $b_{i}^{\dagger}$
  satisfy the generalized quon algebra:

  \begin{equation}
  b_{i}b_{j}^{\dagger} - q_{ij}^{'}b_{j}^{\dagger}b_{i} = \delta_{ij}
  ,\forall i,j \in S .
  \end{equation}

  We have denoted these operators by $b_{i}$ and $b_{i}^{\dagger}$
  because they, in general, differ from the corresponding one in the in-region.
  The general relation between the two sets of in- and out- operators is given
  by

  \begin{equation}
  a_{i} = \sum_{j}(\alpha_{ji}b_{j} + \beta_{ji}^{*}b_{j}^{\dagger}),
  \end{equation}

  where $\alpha_{ji}$, $\beta_{ji}$ are Bogoliubov coefficients for the
  expanding universe \cite{JWG},\cite{LP}. However, for the spacetime with the
  Robertson-Walker metric between the in- and out- regions, $t_{1} < t < t_{2}$
  , the Bogoliubov coefficients are diagonal \cite{LP}. Therefore, we restrict
  ourselves to the case where

  \begin{equation}
  \begin{array}{c}
  \alpha_{ij} = \alpha_{i}\delta_{ij}, \\
  \beta_{ij} = \beta_{i}\delta_{ij}.
  \end{array}
  \end{equation}

  Substituting (4,5) and its complex conjugate into (1), we easily obtain

  \begin{equation}
  \begin{array}{l}
  \alpha_{i}\alpha_{j}^{*}(b_{i}b_{j}^{\dagger} - q_{ij}b_{j}^{\dagger}b_{i})
   + \beta_{i}^{*}\alpha_{j}^{*}(b_{i}^{\dagger}b_{j}^{\dagger} - q_{ij}
   b_{j}^{\dagger}b_{i}^{\dagger}) + \alpha_{i}\beta_{j}(b_{i}b_{j} - q_{ij}
   b_{j}b_{i})\\ + \beta_{i}^{*}\beta_{j}(b_{i}^{\dagger}b_{j} - q_{ij}b_{j}
   b_{i}^{\dagger}) = \delta_{ij} .
  \end{array}
  \end{equation}

  Taking the expectation value of (6) with a vacuum state in the out-region
  yields,

  \begin{equation}
  \mid \alpha_{i}\mid ^{2} - q_{ii}\mid \beta_{i}\mid ^{2} = 1
  , \forall i\in S.
  \end{equation}

  {} From (7) it follows that (6) can be reduced to

  \begin{equation}
  \begin{array}{l}
  \alpha_{i}\beta_{j}(b_{i}b_{j} - q_{ij}b_{j}b_{i}) +
  \beta_{i}^{*}\alpha_{j}^{*}
  (b_{i}^{\dagger}b_{j}^{\dagger} - q_{ij}b_{j}^{\dagger}b_{i}^{\dagger}) +
  (1 - q_{ij}q_{ji}^{'})\beta_{i}^{*}\beta_{j}b_{i}^{\dagger}b_{j} +\\
  (q_{ij}^{'} - q_{ij})\alpha_{i}\alpha_{j}^{*}b_{j}^{\dagger}b_{i} = 0 .
  \end{array}
  \end{equation}

  It is easy to see that the remaining relations following from (8) are

  \begin{equation}
  \begin{array}{c}
  \alpha_{i}\beta_{j}(b_{i}b_{j} - q_{ij}b_{j}b_{i}) = 0
  ,\forall i,j\in S \\
  \beta_{i}^{*}\alpha_{j}^{*}(b_{i}^{\dagger}b_{j}^{\dagger} -
  q_{ij}b_{j}^{\dagger}b_{i}^{\dagger}) = 0    ,\forall i,j\in S \\
  (1 - q_{ij}q_{ij}^{'*})\beta_{i}^{*}\beta_{j} = 0    ,i \neq j \\
  (q_{ij}^{'} - q_{ij})\alpha_{i}\alpha_{j}^{*} = 0    ,i \neq j  \\
  (q_{ii} - q_{ii}^{'})\mid \alpha_{i}\mid ^{2} = (1 - q_{ii}q_{ii}^{'})
  \mid \beta_{i}\mid ^{2}    ,\forall i\in S .
  \end{array}
  \end{equation}

  Now, according to the possible values of Bogoliubov coefficients, we shall
  classify and discuss all possible solutions of the conditions (7), (9). \\
  \\

  Solution (i):   $\alpha_{i}\beta_{i} \neq 0$    ,$\forall i\in S$

  {} From the first of Eqs. (9) it follows that

  \begin{equation}
  b_{i}b_{j} - q_{ij}b_{j}b_{i} = 0    ,\forall i,j\in S
  \end{equation}

  and its hermitian conjugate. These commutation relations between $b_{i}$,
  $b_{j}$
  can exist only if $\mid q_{ij}\mid = 1$ for $\forall i,j\in S$, \cite{SM}.

  {}From Eq. (2) it follows that $q_{ij}^{*} = q_{ij}$. Hence,$q_{ij}^{2} = 1$,
  i.e., $q_{ij} = \pm1$, $\forall i,j\in S$. Furthermore, from the fourth and
  fifth
  of Eqs. (9) one obtains $q_{ij}^{'} = q_{ij}$. Let us summarize solution (i):
  $q_{ij}^{'} = q_{ij}$, $q_{ij}^{2} = 1$ for every pair i,j.\\
  Let us further discuss two possible cases. First, if the index i is a
  continuous
  variable, then the functions $\alpha_{i}$, $\beta_{i}$ are continuous
  functions of i.
  It follows from Eq. (7) that $q_{ii} = const.$, i.e.,  $q_{ii} = +1$,
  $\forall i\in S$ or $q_{ii} = -1$, $\forall i\in S$.
  There are no additional restrictions on $q_{ij}$ if $i \neq j$.
  Note that, in the above considerations, $q_{ij}$ has not necessarily been a
  continuous function. (However, if we assume that $q_{ij}$ is also a
continuous
  function in the variables i,j then $q_{ij} = const. = +1 or -1$ ,$i,j\in S$.)
  Note that $q_{ii}$ would have jumps if $\beta_{i}$ could vanish.\\
  The second possibility assumes that in addition to the continuous variable i
  there is a discrete variable $\mu = 1,....,p$. All the conditions (Eqs.(7),
  (9)) remain the same under the following substitutions: $\alpha_{i},
\beta_{i}
  \rightarrow \alpha_{i}^{\mu}, \beta_{i}^{\mu}$,
  $a_{i}, b_{i} \rightarrow a_{i}^{\mu}, b_{i}^{\mu}$, and $q_{ij}
  \rightarrow q_{i\mu,j\nu}$.
  If we assume that $q_{i\mu,j\nu}$ is a continuous function in the i,j
  variables, then $q_{i\mu,j\nu}$ reduces to $q_{\mu\nu}$.
  Solution (i) reads: $q_{\mu\mu} \neq const.$, $q_{\mu\nu}^{'} = q_{\mu\nu}$
  , $q_{\mu\nu}^{2} = 1$, $\mu,\nu = 1,...,p$.\\
  Let us write the corresponding algebra:

  \begin{equation}
  \begin{array}{c}
  a_{i}^{\mu}a_{j}^{\nu\dagger} - q_{\mu\nu}a_{j}^{\nu\dagger}a_{i}^{\mu} =
  \delta_{ij}\delta_{\mu\nu}    ,\forall \mu,\nu = 1,...,p\\
  a_{i}^{\mu}a_{j}^{\nu} - q_{\mu\nu}a_{j}^{\nu}a_{i}^{\mu} = 0
  ,\forall i,j\in S .
  \end{array}
  \end{equation}

  This is a generalization of Green's oscillator algebra \cite{HSG}. In the
  special case $q_{\alpha\beta} = \pm(2\delta_{\alpha\beta} - 1)$, one obtains
  p Green's oscillators of the Bose (Fermi) type for the upper (lower) sign.
  Hence, para-Bose and para-Fermi statistics are also allowed, in addition to
  the ordinary Bose and Fermi statistics.

  Solution (ii):    $\alpha_{i} \neq 0$, $\beta_{i} = 0$   ,$\forall i\in S$

  Equation (7) gives $\mid \alpha_{i}\mid = 1$, $i\in S$ and the fourth and the
  fifth of Eqs. (9) give $q_{ij}^{'} = q_{ij}$. In this case there are no
  restrictions on statistics. We can have arbitrary $q_{ij}$, with $\mid q_{ij}
  \mid \leq 1$
  and $q_{ij}^{'} = q_{ij}$. In this case anyons in a multivalued picture
  \cite{BDM}
  are also allowed.

  Solution (iii):   $\alpha_{i} = 0$, $\beta_{i} \neq 0$    ,$\forall i\in S$

  {} From Eq. (7) it follows that $-q_{ii}\mid \beta_{i}\mid ^{2} = 1$ and
  $q_{ii} < 0$, $\forall i\in S$.
  Since $\beta_{i}\beta{j}^{*} \neq 0$, the third of Eqs. (9) gives
  $q_{ij}^{'*} = q_{ij}^{-1}$.
  In addition, $\mid q_{ij}^{'}\mid \leq 1$ and $\mid q_{ij}\mid \leq 1$,
  therefore $\mid q_{ij}^{'}\mid = \mid q_{ij}\mid = 1$. Hence, we find that
  $q_{ij}^{'} = q_{ij} = e^{i\triangle_{ij}}$, $i \neq j$, where
  $\triangle_{ij} = -\triangle_{ji}$ is a real function.
  For $i = j$, $q_{ii}^{'} = q_{ii}^{-1}$, and in the same way as above we find
  that $q_{ii}^{'} = q_{ii} = \pm1$. Since $q_{ii} < 0$, the only possibility
  is $q_{ii}^{'} = q_{ii} = -1$, which means the hard-core condition for all
  oscillators: $a_{i}^{2} = b_{i}^{2} = 0$, $\forall i\in S$.\\
  Let us summarize solution (iii):\\
  $q_{ij}^{'} = q_{ij} = e^{i\triangle_{ij}}$, $i \neq j$ and $q_{ii}^{'} =
  q_{ii} = -1$, $\forall i\in S$.
  It corresponds to anyons with the hard-core condition in a single-valued
  picture \cite{LS}.
  \\

  General solution (iv):

  Solutions (i), (ii) and (iii) are characterized by the respective restriction
  s: $\alpha_{i}\beta_{i} \neq 0$, $\forall i\in S$;  $\alpha_{i} \neq 0$,
  $\beta_{i} = 0$, $\forall i\in S$;
  $\alpha_{i} = 0$, $\beta_{i} \neq 0$, $\forall i\in S$. According to
  \cite{JWG},\cite{LP}, solution (ii)
  is not physically acceptable. In a similar way, solution (iii) seems to be
  not allowed
  unless there is a special dynamical reason for that. So we are left with
  solution (i)
  allowing the Bose, Fermi, and generalized Green's oscillator algebra to be
  consistent
  with the dynamical evolution of spacetime.\\
  However, even solution (i), might not be realistic. Therefore, we now assume
  no restrictions on $\alpha_{i}$, $\beta_{i}$, except the necessary condition
  Eq.(7).
  Let us decompose the initial set of indices i into three not overlapping
  subsets:

  \begin{equation}
  \begin{array}{c}
  M_{1} = \{ i\in S, \alpha_{i}\beta_{i} \neq 0 \},\\
  M_{2} = \{ i\in S, \beta_{i} = 0 \},\\
  M_{3} = \{i\in S, \alpha_{i} = 0 \}.
  \end{array}
  \end{equation}

  Note that $M_{1} \bigcup M_{2} \bigcup M{3} = S$.\\
  It is obvious that solution (1) applies to set $M_{1}$, solution (ii) to set
  $M_{2}$,
  and solution (iii) to set $M_{3}$.
  Equations (9) with indices i,j from different subsets ,($M_{1}$,$M_{2}$),
  ($M_{1}$,$M_{3}$), ($M_{2}$,$M_{3}$), are satisfied by $q_{ij}^{'} =
  q_{ij} = \pm1$.
  Hence the general solution for $\alpha_{i}$, $\beta_{i}$, calculated in a
  dynamical model of evolving spacetime, allows all types of statistics,
  at least in principle. The main conclusion is that $q_{ij}^{'} = q_{ij}$,
i.e.,
  that $q_{ij}$ cannot be changed in the course of dynamical evolution.
  The question of existence, structure, and topology of subsets $M_{2}$,
$M_{3}$
  is deeply connected with a concrete dynamical model of evolving spacetime,
and
  is beyond the scope of this paper.\\
  Nevertheless, let us assume that nontrivial sets $M_{2}$, $M_{3}$ exist.
  Hence, there may exist several different regions (i.e., "phases" in momentum
  space)
  with solutions $M_{1}$ of the Bose or Fermi type. The regions are separated
  by a boundary of the type $M_{2}$ or $M_{3}$. All types of statistics
  (including
  infinite-quonic statistics) are possible on the $M_{2}^{\beta=0}$ boundary.
  Only anyons with the hard-core condition are possible on the
$M_{3}^{\beta=0}$
  boundary. Transition from one $M_{1}$ phase to another $M_{1}$ phase through
  $M_{3}^{\beta=0}$ may change the Bose-type to Fermi-type of statistics.
  However, transition through $M_{3}^{\alpha=0}$ is possible only if the
  neighboring phases are of the Fermi-type and, hence, cannot change the type
of
  statistics.\\
  Finally, let us mention that our analysis, although general, is still
  restricted
  to the generalized quon algebra (Eq.(1)). However, there are other deformed
  algebras \cite{PW}, not included in Eq.(1), for which a similar analysis can
  be performed.
  \\
  \\

  This work was supported by the Scientific Fund of the Republic of Croatia.

  %******************REFERENCES**********************************************
  
  \end{document}